\documentclass[prl,preprint,showpacs,amsmath,amssymb,superscriptaddress,floatfix]{revtex4-1}
\usepackage{dcolumn}
\usepackage{color}
\usepackage[dvips]{graphicx}
\usepackage{amsmath}
\usepackage{multirow}
\usepackage{bm}

\newcommand\T{\rule{0pt}{3.1ex}}

\begin{document}

\title{Limits on the neutrino mass from neutrinoless double-$\beta $ decay}

\author{J.\ Barea}
\email{jbarea@udec.cl}
\affiliation{Departamento de F\'{i}sica, Universidad de Concepci\'{o}n,
 Casilla 160-C, Concepci\'{o}n, Chile}

\author{J. Kotila}
\email{jenni.kotila@yale.edu}
\affiliation{Center for Theoretical Physics, Sloane Physics Laboratory,
 Yale University, New Haven, CT 06520-8120, USA}

\author{F.\ Iachello}
\email{francesco.iachello@yale.edu}
\affiliation{Center for Theoretical Physics, Sloane Physics Laboratory,
 Yale University, New Haven, CT 06520-8120, USA}

\begin{abstract}
Neutrinoless double-$\beta$ decay is of fundamental importance for the determining neutrino mass. By combining a calculation of nuclear matrix elements
within the framework of the microscopic interacting boson model (IBM-2) with
an improved calculation of phase space factors, we set limits on the average
light neutrino mass and on the average inverse heavy neutrino mass (flavor
violating parameter).
\end{abstract}


\maketitle

The process $0\nu \beta \beta $ in which a nucleus X is transformed into a
nucleus Y with the emission of two electrons and no neutrinos, $_{Z}^{A}$X$%
_{N}\rightarrow $ $_{Z+2}^{A}$Y$_{N-2}$+2e$^{-}$, is of fundamental
importance for determining the Majorana or Dirac nature of the neutrino and confirming a non-zero value of its mass as established by neutrino oscillation experiments \cite{fukuda, ahmad, eguchi},  what constitutes physics beyond the standard model. The half-life for this process
can be written as 
\begin{equation}
\left[ \tau _{1/2}^{(0\nu )}\right] ^{-1}=G_{0\nu }\left\vert M_{0\nu
}\right\vert ^{2}\left\vert f(m_{i},U_{ei})\right\vert ^{2},
\end{equation}%
where $G_{0\nu }$ is a phase space factor (PSF), $M_{0\nu }$ is the nuclear
matrix element (NME), and $f$ contains physics beyond the standard model
through the masses $m_{i}$ and elements $U_{ei}$ of the mixing matrix of the
neutrino (or other hypothetical particle beyond the standard model). We have
recently (i) introduced a new method \cite{barea}, the microscopic
interacting boson model, IBM-2, to calculate the NME in a consistent way for
all nuclei of interest, and (ii) improved the calculation of the phase space
factors (PSF) by solving the Dirac equation for the outgoing electrons in
the presence of a charge distribution and including electron screening \cite%
{kotila}. In this letter, we present results of a calculation that combines
the NMEs and the PSFs to half-lives. By comparing with current experimental
limits we then set limits on neutrino masses and their couplings.

Starting from the weak Lagrangean, $\mathcal{L}$, one can derive the
transition operator inducing the decay, which, under certain circumstances,
can be factorized as $T(p)=H(p)f(m_{i},U_{ei})$, where $p=\left\vert \vec{q}%
\right\vert $ is the momentum transferred to the leptons \cite{doi, tomoda,simkovic}. The transition operator $H(p)$ has the form 
\begin{equation}
H(p)=\tau _{n}^{\dag }\tau _{n^{\prime }}^{\dag }\left[ -h^{F}(p) +h^{GT}(p)%
\vec{\sigma}_{n}\cdot \vec{\sigma}_{n^{\prime }} \right.
\left. +h^{T}(p)S_{nn^{\prime }}^{p}%
\right] .
\end{equation}%
The factors $h^{F,GT,T}(p)$ are given by $h^{F,GT,T}(p)=v(p)\tilde{h}%
^{F,GT,T}(p)$, where $v(p)$ is called the neutrino "potential" and $\tilde{h}%
(p)$ are the form factors, listed in Ref. \cite{simkovic}. This form assumes the
closure approximation which is expected to be good a approximation for $0\nu \beta \beta $ decay \cite{suh91, suh98}
since the neutrino momentum is of the order of $100$ MeV/c while the energy
scale of the nuclear excitations is $1$ MeV, and all multipoles in the intermediate odd-odd nucleus contribute to the decay. (Conversely, the approximation is not expected to be good  for $2\nu\beta\beta$ decay, where the neutrino momentum is of order $2$ MeV/c, and only $1^+$ and $0^+$ states in the intermediate odd-odd nucleus contribute to the decay). The finite nucleon size is
taken into account by taking the coupling constants momentum dependent and
short range correlations (SRC) are taken into account by convoluting $v(p)$
with the correlation function $J(p)$ taken as a Jastrow function. The
functions $f(m_{i},U_{ei})$ and $H(p)$ depend on the model of $0\nu \beta
\beta $ decay. We consider here explicitly two cases: (i) the emission and
reabsorption of a light ($m_{light}\ll 1$ keV) neutrino; (ii) the emission
and reabsorption of a heavy ($m_{heavy}\gg 1$ GeV) neutrino. For scenario
(i), the function $f$ can be written as%
\begin{equation}
f=\frac{\left\langle m_{\nu }\right\rangle }{m_{e}},\text{ \ \ \ \ \ \ }%
\left\langle m_{\nu }\right\rangle =\sum_{k=light}\left( U_{ek}\right)
^{2}m_{k},
\end{equation}%
where $U$ is the neutrino mixing matrix. The average neutrino mass is given
in terms of mixing angles and phases \cite{fogli} and is constrained by
atmospheric, solar and neutrino oscillation experiments. The potential $v(p)$
for this case is $v(p)=2\pi^{-1}[p(p+\tilde{A})]^{-1}$ where $\tilde{%
A}$ is the so-called closure energy. For scenario (ii) the transition
operator can be written as $T_{h}(p)=H_{h}(p)f_{h}(m_{i},U_{ei})$, where the
index $h$ refers to heavy. The function $f_{h}$ can be written as 
\begin{equation}
f_{h}=m_{p}\left\langle \frac{1}{m_{h}}\right\rangle ,\text{ \ \ }%
\left\langle \frac{1}{m_{h}}\right\rangle =\sum_{k=heavy}\left(
U_{ek_{h}}\right) ^{2}\frac{1}{m_{k_{h}}}.
\end{equation}%
The neutrino potential is $v_{h}(p)=2\pi^{-1 }(m_{e}m_{p})^{-1}$. The
function $f_{h}$ is often written as $\eta $ and called the flavor violating
parameter. The average inverse heavy neutrino mass has in the past been
considered as an unconstrained parameter. 
However, recently, it has been
suggested \cite{vissani} that some constraints can be put on this quantity
from
 large hadron collider (LHC) physics and lepton flavor violating processes.
 The effect of heavy
neutrinos on neutrinoless double-$\beta$ decay has been illustrated within the
context of a specific model as a function of the mass of the lightest heavy
neutrino in the range $1$-$500$ GeV.

\begin{table}
 \caption{\label{table1}Neutrinoless double-$\beta$ decay matrix elements $M^{(0\nu )}$ in IBM-2 with Argonne CCM SRC and $g_A=1.269$, in QRPA with Argonne CCM SRC and $g_A=1.254$, and ISM with UCOM SRC and $g_A=1.25$.} 
 \begin{ruledtabular} %
\begin{tabular}{cccc}
A		&IBM-2     &QRPA\footnotemark[1]    &ISM\footnotemark[2]  \\
\hline 
\T
48		& 2.28 	&						&0.85\tabularnewline
76   	& 5.98 	&5.81					&2.81\tabularnewline
82    	& 4.84 	&5.19					&2.64\tabularnewline
96    	& 2.89 	&1.90					&\tabularnewline
100   	& 4.31 	&4.75					& \tabularnewline
110    	& 4.15 	&						& \tabularnewline
116    	& 3.16 	&3.54					& \tabularnewline
124    	& 3.89 	&						&2.62 \tabularnewline
128    	& 4.97	&4.93					&2.88 \tabularnewline
130    	& 4.47 	&4.37					&2.65 \tabularnewline
136    	& 3.67 	&2.78					&2.19 \tabularnewline
148    	& 2.36 	&						& \tabularnewline
150   	& 2.74 	&						& \tabularnewline
154    	& 2.91 	&						& \tabularnewline
160    	& 4.17 	&						& \tabularnewline
198    	& 2.25 	&						& \tabularnewline
\end{tabular}\end{ruledtabular} 
\footnotetext[1]{Ref.~\cite{simkovic1}}
\footnotetext[2]{Ref.~\cite{menendez}}
\end{table}
\begin{figure}[cbt!]
\begin{center}
\includegraphics[width=8.6cm]{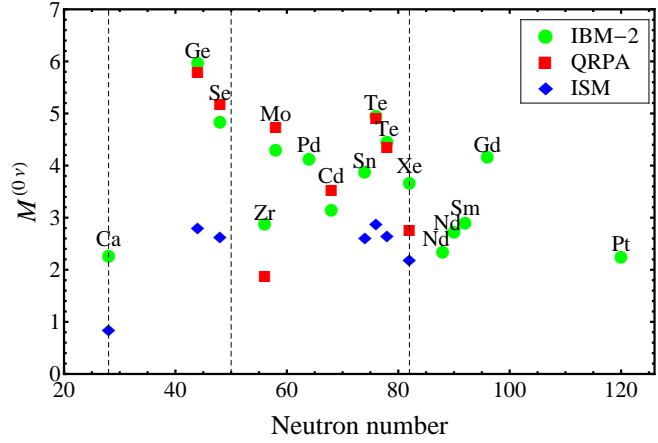} 
\end{center}
\caption{\label{fig1}(Color online) Nuclear matrix elements $M^{(0\nu )}$
for $0\nu \beta \beta$ decay in IBM-2 compared with QRPA \cite{simkovic1} and ISM \cite{menendez}.}
\end{figure}
\begin{table}
 \caption{\label{table2} Neutrinoless double-$\beta$ decay matrix elements $M_{h}^{(0\nu )}$ in IBM-2 with Argonne CCM SRC and $g_A=1.269$, and in QRPA with Argonne CCM SRC, $g_A=1.25$ and intermediate size for the model space. }
 \begin{ruledtabular} %
\begin{tabular}{ccc}
$A$  &IBM-2 &QRPA\footnotemark[1] \tabularnewline
\hline 
\T
48  	&46.3	&		  \tabularnewline
76  	&107	&233   	  \tabularnewline
82  	&84.4	&226   	  \tabularnewline
96		&99.0	&   	  \tabularnewline
100 	&165	&250   	  \tabularnewline
110 	&155	&   	  \tabularnewline
116 	&110.	&	  	  \tabularnewline
124 	&79.6	&   	  \tabularnewline
128 	&101	&   	  \tabularnewline
130 	&92.0	&234   	  \tabularnewline
136 	&72.8		&   	  \tabularnewline
148 	&103	&   	  \tabularnewline
150 	&116	&  	  \tabularnewline
154 	&113	&   	  \tabularnewline
160 	&155	&   	  \tabularnewline
198 	&104	&   	  \tabularnewline
\end{tabular}\end{ruledtabular} 
\footnotetext[1]{Ref.~\cite{faessler2011} }
\end{table}
\begin{ruledtabular}
\begin{table}[cbt!]
\caption{\label{table3}Left: Calculated half-lives in IBM-2 for neutrinoless double-$\beta$ decay for $\left<m_{\nu}\right>=1$ eV and $g_A=1.269$. Right: Upper limit on neutrino mass from current experimental limit from a compilation of Barabash \cite{barabash11}. The value reported by Klapdor-Kleingrothaus \textit{et al.} \cite{klapdor}, the limit from IGEX \cite{igex}, and the recent limits from KamLAND-Zen \cite{kamland} and EXO \cite{exo} are also included.}
\begin{tabular}{lc|cc}
Decay  &  \ensuremath{\tau_{1/2}^{0\nu}}(\ensuremath{10^{24}}yr) &  \ensuremath{\tau_{1/2, exp}^{0\nu}}(yr) &$\left< m_{\nu}\right>$(eV)\\
 \hline
 \T
$^{48}$Ca$\rightarrow ^{48}$Ti		&0.782 &$>5.8\times 10^{22}$ &$<3.7$\\
$^{76}$Ge$\rightarrow ^{76}$Se	 	&1.19 &$>1.9\times 10^{25}$ &$<0.25$\\
								 	&	 	&$1.2\times 10^{25}$\footnotemark[1] &$0.32$\\
								 	&	 	&$>1.6\times 10^{25}$\footnotemark[2] &$<0.27$\\

$^{82}$Se$\rightarrow ^{82}$Kr	 	&0.423 &$>3.6\times 10^{23}$ &$<1.1$\\
$^{96}$Zr$\rightarrow ^{96}$Mo		&0.588 &$>9.2\times 10^{21}$ &$<8.0$\\
$^{100}$Mo$\rightarrow ^{100}$Ru 	&0.340 &$>1.1\times 10^{24}$ &$<0.56$\\
$^{110}$Pd$\rightarrow ^{110}$Cd 	&1.22 & &\\
$^{116}$Cd$\rightarrow ^{116}$Sn  	&0.602 &$>1.7\times 10^{23}$ &$<1.9$\\
$^{124}$Sn$\rightarrow ^{124}$Te 	&0.737 & &\\
$^{128}$Te$\rightarrow ^{128}$Xe 	&6.94 &$>1.5\times 10^{24}$ &$<2.2$\\
$^{130}$Te$\rightarrow ^{130}$Xe 	&0.355 &$>2.8\times 10^{24}$ &$<0.36$\\
$^{136}$Xe$\rightarrow ^{136}$Ba 	&0.512 &$>5.7\times 10^{24}$\footnotemark[3] &$<0.30$\\
								 	&	 	&$>1.6\times 10^{25}$\footnotemark[4] &$<0.18$\\

$^{148}$Nd$\rightarrow ^{148}$Sm 	&1.79 & &\\
$^{150}$Nd$\rightarrow ^{150}$Sm 	&0.213 &$>1.8\times 10^{22}$ &$<3.4$\\
$^{154}$Sm$\rightarrow ^{154}$Gd 	&3.94 & &\\
$^{160}$Gd$\rightarrow ^{160}$Dy 	&0.606 & &\\
$^{198}$Pt$\rightarrow ^{198}$Hg 	&2.64 & &\\
\end{tabular}
\footnotetext[1]{Ref.~\cite{klapdor}}
\footnotetext[2]{Ref.~\cite{igex}}
\footnotetext[3]{Ref.~\cite{kamland}}
\footnotetext[4]{Ref.~\cite{exo}}
\end{table}
\end{ruledtabular}

\begin{figure}[cbt!]
\begin{center}
\includegraphics[width=8.6cm]{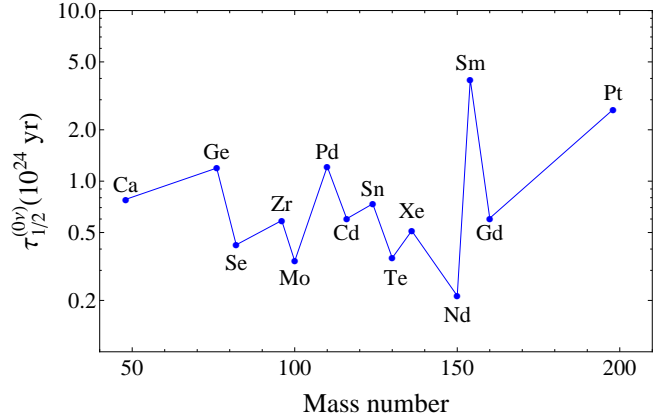} 
\end{center}
\caption{\label{fig2}(Color online) Expected half-lives for $\left\langle m_{\nu}\right\rangle=1$ eV, $g_{A}=1.269$. The points for $^{128}$Te and $^{148}$Nd decays are not included in this figure. The figure is in semilogarithmic scale.}
\end{figure}

We have calculated the nuclear matrix elements within the framework of the
microscopic interacting boson model, IBM-2 \cite{otsuka}, in all nuclei of
interest. Details of the calculation are given in Ref. \cite{barea} and in a
forthcoming long publication \cite{bki}. Matrix elements $M^{(0\nu )}$ for
light neutrino exchange are shown in\ Table~\ref{table1} and Fig.~\ref{fig1}, where they are
compared with those calculated with other methods, most notably QRPA \cite%
{simkovic1} and ISM 
\cite{menendez} with the same (or similar)
approximations for the SRC. We note both in Table~\ref{table1} and Fig.~\ref{fig1} a close
correspondence between the IBM-2 and QRPA calculations, while the ISM
results are approximately a factor of 2 smaller than IBM-2/QRPA. 
(The origin of the difference is not completely clear. The three models make different approximations and at different levels. A recent combined analysis of $0\nu\beta\beta$ and $2\nu\beta\beta$ decay \cite{bki} seems to indicate that the main difference is the size of the model space in which the calculations are done. This is substantiated by the observation that the behavior with mass number of all three calculations is similar and that they can be reconciled by a simple renormalization).
Matrix elements $M_{h}^{(0\nu )}$ for heavy neutrino exchange are shown in
Table~\ref{table2}.  By combining the matrix elements with the phase space factors of Ref. \cite%
{kotila}, we obtain the
 expected half-lives shown in Table~\ref{table3}, left, and
Fig.~\ref{fig2} for light neutrino exchange and Table~\ref{table4}, left, for heavy neutrino
exchange. It should be noted that the combination must be done consistently. If the phase space factors of Ref. \cite{kotila} are used, 
the nuclear matrix elements $M^{(0\nu )}$ of Tables~\ref{table1} and ~\ref{table2} must be multiplied by 
$g_{A}^{2}$, that~is~$M_{0\nu }=g_{A}^{2}M^{(0\nu )}$ in Eq. (1).
\begin{ruledtabular}
\begin{center}
\begin{table*}[cbt!]
\caption{\label{table4}Left: Calculated half-lives for neutrinoless double 
$\beta$ decay with exchange of heavy neutrinos for $\eta=2.75\times10^{-7}$ and $g_A=1.269$. Right: Upper limits of $|\eta|$ and lower limits of heavy neutrino mass from current experimental limit from a compilation of Barabash \cite{barabash11}. The value reported by Klapdor-Kleingrothaus \textit{et al.} \cite{klapdor}, the limit from IGEX \cite{igex},  and the recent limits from KamLAND-Zen \cite{kamland} and EXO \cite{exo} are also included.}
\begin{tabular}{lc|ccc}
Decay  &  \ensuremath{\tau_{1/2}^{0\nu_h}}(\ensuremath{10^{24}}yr) &  \ensuremath{\tau_{1/2, exp}^{0\nu_h}}(yr) &$|\eta|(10^{-7})$ &$\left< m_{\nu_h}\right>$(GeV)\\
 \hline
 \T
$^{48}$Ca$\rightarrow ^{48}$Ti		&0.096		&$>5.8\times 10^{22}$	&$<3.54$	&$>0.73$\\
$^{76}$Ge$\rightarrow ^{76}$Se	 	&0.190		&$>1.9\times 10^{25}$	&$<0.275$	&$>9.4$\\
								 	&	 		&$1.2\times 10^{25}$\footnotemark[1]	&$0.346$ &$7.5$\\
								 	&	 		&$>1.6\times 10^{25}$\footnotemark[2]	&$<0.300$ &$>8.6$\\
								 	
$^{82}$Se$\rightarrow ^{82}$Kr	 	&0.070		&$>3.6\times 10^{23}$	&$<1.22$	&$>2.1$\\
$^{96}$Zr$\rightarrow ^{96}$Mo		&0.025		&$>9.2\times 10^{21}$	&$<4.56$	&$>0.6$\\
$^{100}$Mo$\rightarrow ^{100}$Ru 	&0.012		&$>1.1\times 10^{24}$	&$<0.285$	&$>9.1$\\
$^{110}$Pd$\rightarrow ^{110}$Cd 	&0.044		&						&	&\\
$^{116}$Cd$\rightarrow ^{116}$Sn  	&0.025		&$>1.7\times 10^{23}$	&$<1.06$	&$>2.5$\\
$^{124}$Sn$\rightarrow ^{124}$Te 	&0.089		&				&&\\
$^{128}$Te$\rightarrow ^{128}$Xe 	&0.846		&$>1.5\times 10^{24}$	&$<2.07$	&$>1.2$\\
$^{130}$Te$\rightarrow ^{130}$Xe 	&0.042		&$>2.8\times 10^{24}$	&$<3.38$	&$>7.6$\\
$^{136}$Xe$\rightarrow ^{136}$Ba 	&0.066		&$>5.7\times 10^{24}$\footnotemark[3]	&$<0.296$	&$>8.7$\\
								 	&	 		&$>1.6\times 10^{25}$\footnotemark[4]	&$<0.177$ &$>14.6$\\

$^{148}$Nd$\rightarrow ^{148}$Sm 	&0.048		&					&	&\\
$^{150}$Nd$\rightarrow ^{150}$Sm 	&0.006		&$>1.8\times 10^{22}$	&$<1.58$	&$>1.6$\\
$^{154}$Sm$\rightarrow ^{154}$Gd 	&0.132		&	&	&\\
$^{160}$Gd$\rightarrow ^{160}$Dy 	&0.022		&	&	&\\
$^{198}$Pt$\rightarrow ^{198}$Hg 	&0.063		&	&	&\\
\end{tabular}
\footnotetext[1]{Ref.~\cite{klapdor}}
\footnotetext[2]{Ref.~\cite{igex}}
\footnotetext[3]{Ref.~\cite{kamland}}
\footnotetext[4]{Ref.~\cite{exo}}
\end{table*}
\end{center}
\end{ruledtabular}

Using the experimental upper limits from a compilation of Barabash \cite%
{barabash11}, the IBM-2 matrix elements of Tables ~\ref{table1} and ~\ref{table2} and the phase space
factors of \cite{kotila}, we estimate current limits on the neutrino mass
given in Tables~\ref{table3}, right, and Table~\ref{table4}, right, which are the main results
of this letter. In Table~\ref{table4} we give limits both on the flavor violating
parameter $\eta $ and on the average heavy neutrino mass, defined as $
\langle m_{\nu _{h}}\rangle / m_{p}=(M_{W}^{4}/M_{WR}^{4})\eta ^{-1}$, where $M_{W}=80.41\pm 0.10$ GeV and $M_{WR}$ is assumed to be $M_{WR}=3.5$ TeV. While
the former is model independent, the latter depends on the model of
left-right mixing \cite{vissani}.

\begin{figure}[cbt!]
\begin{center}
\includegraphics[width=8.6cm]{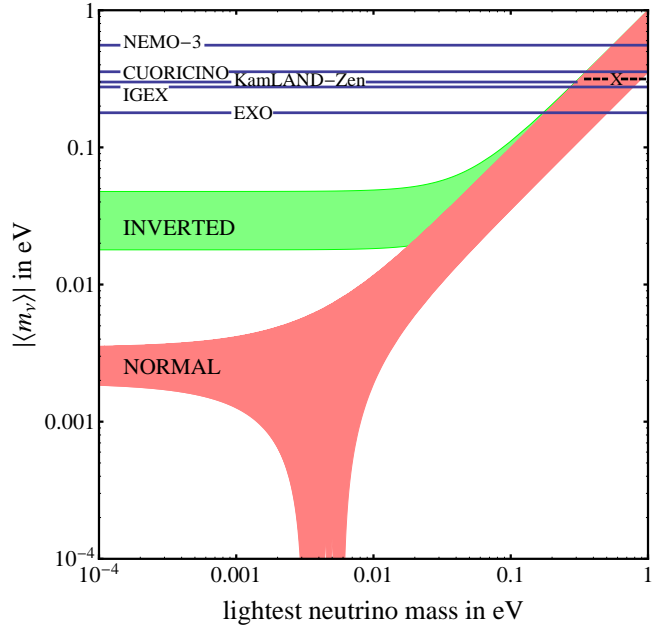} 
\end{center}
\caption{\label{fig3}(Color online) Current limits to $\left\langle m_{\nu}\right\rangle$ from CUORICINO \cite{cuoricino}, IGEX \cite{igex}, NEMO-3 \cite{nemo}, KamLAND-Zen \cite{kamland}, and EXO \cite{exo} and IBM-2 nuclear matrix elements. The value of Ref.~\cite{klapdor} is shown by $X$. It is consistent only with nearly degenerate neutrino masses.  The figure is in logarithmic scale.}
\end{figure}

These results are obtained using the free value of the axial vector coupling
constant as obtained from neutron decay, $g_{A}=1.269$. It is known from
single $\beta $ decay and $2\nu \beta \beta $ decay that $g_{A}$ is
renormalized in nuclei. There are two reasons for the renormalization: (i)
the limited model space within which the calculation of the NME is done;
(ii) the omission of non-nucleonic degrees of freedom ($\Delta ,N^{\ast },$%
...). Since the coupling constant $g_{A}$ appears to the fourth power in the
life-time, the renormalization effect is non negligible and it will amount
to a multiplication of the limits in Table~\ref{table3} and ~\ref{table4} by a factor of 2-4.
Details of the renormalization procedure, as well as of the calculation of
the renormalized matrix elements NME, will be given in a forthcoming longer
publication \cite{bki}. The question of whether or not $0\nu \beta \beta $
matrix elements should be renormalized as much as $2\nu \beta \beta $ matrix
elements is the subject of much debate. In $2\nu \beta \beta $ only 1$^{+}$
and 0$^{+}$ states in the intermediate odd-odd nucleus contribute to the
decay, while in $0\nu \beta \beta $ all multipoles play a role. In this
letter we do not dwell on this question, but rather present results with the
unrenormalized value $g_{A}=1.269$, summarized in\ Fig.~\ref{fig3}. 
From this figure, one can see that in the immediate future only the
degenerate region can be tested by experiments and that the exploration of
the inverted region must await much
larger ($>1$ton) experiments, especially if $g_{A}$ in $0\nu \beta \beta $
is renormalized as much as in $2\nu \beta \beta $ decay. From the same figure, one can also see that even the one-ton experiments will not be able to reach into the normal hierarchy.

This work was performed in part under the USDOE Grant DE-FG02-91ER-40608 and
Fondecyt Grant No. 1120462.

\end{document}